\documentclass[conference]{IEEEtran}
\IEEEoverridecommandlockouts
\usepackage{cite}
\usepackage{amsmath,amssymb,amsfonts}
\usepackage{algorithmic}
\usepackage{graphicx}
\usepackage{textcomp}
\usepackage{xcolor}

\def\BibTeX{{\rm B\kern-.05em{\sc i\kern-.025em b}\kern-.08em
    T\kern-.1667em\lower.7ex\hbox{E}\kern-.125emX}}
\usepackage{import}
\usepackage{amsmath}
\usepackage{comment}
\usepackage{xspace}

\usepackage{enumitem}
\usepackage{amssymb}
\usepackage{graphicx}
 \usepackage{xcolor}
\usepackage[linesnumbered,ruled,vlined,algo2e]{algorithm2e}
\usepackage{hyperref}
\hypersetup{
    colorlinks=true,
    linkcolor=blue,
    filecolor=black,      
    urlcolor=cyan,
}
\usepackage[format=default,font=small,labelfont=bf]{caption}
\usepackage{changepage}
\usepackage{adjustbox}
\usepackage{array}
\newcolumntype{P}[1]{>{\centering\arraybackslash}p{#1}}
\newcolumntype{R}[1]{>{\raggedleft\arraybackslash}p{#1}}

\def\fullversion{0}

\newtheorem{definition}{Definition}
\newtheorem{theorem}{Theorem}

\makeatletter
\newcommand*{\rom}[1]{\expandafter\@slowromancap\romannumeral #1@}
\makeatother

\newcommand{\revision}[1]{#1}

\newcommand{\framework}{\textsc{NCExplorer}\xspace}
\newcommand{\lucene}{\textsc{Lucene}\xspace}
\newcommand{\bert}{\textsc{Bert}\xspace}
\newcommand{\newslink}{\textsc{Newslink}\xspace}
\newcommand{\newsbert}{\textsc{Newslink-Bert}\xspace}

\newcommand{\rollup}{\emph{roll-up}\xspace}
\newcommand{\drilldown}{\emph{drill-down}\xspace}

\newcommand{\vsetins}{\mathcal{V}_{I}}
\newcommand{\vsetcon}{\mathcal{V}_{C}}
\newcommand{\esetins}{\mathcal{E}_{I}}
\newcommand{\esetcon}{\mathcal{E}_{C}}
\newcommand{\onto}{\Psi}

\newcommand{\kgentity}[1]{``\textit{#1}''}

\begin{document}



\title{Enabling Roll-up and Drill-down Operations in News Exploration with Knowledge Graphs for Due Diligence and Risk Management \thanks{Preprint. Accepted @ ICDE 2024. This research is supported by the Lee Kong Chian fellowship. Zhifeng Bao is supported in part by DP240101211.}}

%


\author{
    \IEEEauthorblockN{Sha Wang\(^{1}\), Yuchen Li\(^{1}\), Hanhua Xiao\(^1\), Zhifeng Bao\(^2\), Lambert Deng\(^3\) Yanfei Dong\(^4\)
    }
    \IEEEauthorblockA{
    \(^1\) Singapore Management University  \(^2\) RMIT University
    \(^3\) DBS Bank Limited 
    \(^4\) PayPal, National University of Singapore\\
    \texttt{sha.wang.2021@phdcs.smu.edu.sg, yuchenli@smu.edu.sg, hhxiao.2020@phdcs.smu.edu.sg}\\
    \texttt{zhifeng.bao@rmit.edu.au, lambertdeng@dbs.com, dyanfei@paypal.com}
    }
}

\maketitle

\begin{abstract}
Efficient news exploration is crucial in real-world applications, particularly within the financial sector, where numerous control and risk assessment tasks rely on the analysis of public news reports. The current processes in this domain predominantly rely on manual efforts, often involving keyword-based searches and the compilation of extensive keyword lists. In this paper, we introduce \framework, a framework designed with OLAP-like operations to enhance the news exploration experience. \framework empowers users to use \rollup operations for a broader content overview and \drilldown operations for detailed insights. These operations are achieved through integration with external knowledge graphs (KGs), encompassing both fact-based and ontology-based structures. This integration significantly augments exploration capabilities, offering a more comprehensive and efficient approach to unveiling the underlying structures and nuances embedded in news content.
%
Extensive empirical studies through master-qualified evaluators on Amazon Mechanical Turk demonstrate \framework's superiority over existing state-of-the-art news search methodologies across an array of topic domains, using real-world news datasets. 


\end{abstract}


\section{Introduction}
\label{chap:introduction}

\begin{figure*}[t]
  
      \caption{\framework \textbf{\rollup} and \textbf{\drilldown} example. A user can roll-up a specific entity \kgentity{FTX} to an abstract topic\kgentity{Bitcoin Exchange} and drill-down to other topics in the result like \kgentity{Regulator}. Yellow boxes indicate user operations and green boxes show results.}
    \vspace{-0.5em}

    \includegraphics[width=\linewidth]{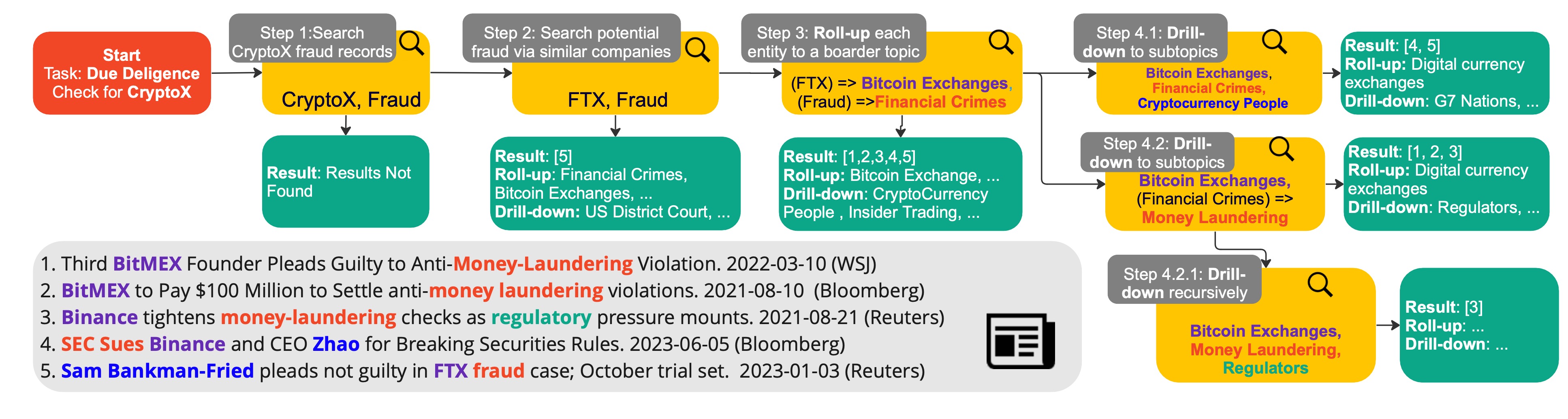}
    \label{fig:rollup-drilldown-example}
    \vspace{-3.5em}
\end{figure*}

Risk assessment and due diligence are paramount for financial institutions, merchants, and all relevant stakeholders.
For instance, global payment giant PayPal conducts Anti-Money Laundering (AML) and Counter-Terrorist Financing (CTF) risk assessments, as well as sanctions screening, for every new business client~\cite{compliance-commitment-paypal}. Similarly, DBS, Southeast Asia's largest bank, shoulders additional responsibilities in the realm of Environmental, Social, and Governance (ESG) concerns~\cite{dbs-responsible-finance}. DBS's ESG risk policy explicitly prohibits financial involvement in activities associated with illegal logging, forced or child labor, wildlife trading, and more. 
In risk assessments, analysts in financial institutions rely on public news reports to determine whether the entities under scrutiny have ever been associated with suspicious activities. Due to the complexity of these due diligence tasks, they are predominantly manual. Compliance teams laboriously maintain extensive lists of financial crime terminology and sift through search results to distinguish genuine financial misconduct from unrelated or benign news. This creates a big operational overhead of running a business. According to a recent McKinsey survey~\cite{compliance-cost-mckinsey}, compliance-related expenses in major banks have surged, becoming unsustainable. The situation becomes even more precarious for DNFBPs (Designated Non-Financial Businesses and Professions), such as precious metal dealers or real estate agents, often operating with limited resources. There is a clear need for a more efficient approach to news exploration. 

In our research, we introduce an OLAP-inspired approach to news exploration. In the \rollup process, users input known terms, leading to the generation of broader topics. For example, \kgentity{FTX} is expanded to \kgentity{Bitcoin Exchange}. 
Our system, \framework, subsequently amplifies these topics by curating a list of relevant keywords for retrieval. The retrieved articles come with an array of related subtopics, granting users the capability to \drilldown into specific news pieces and discover unanticipated topics. 

To illustrate the enhanced productivity gained with our approach, especially in due diligence checks,  consider a Know Your Customer (KYC) task involving a newly incorporated cryptocurrency exchange, \kgentity{CryptoX}, as depicted in Fig \ref{fig:rollup-drilldown-example}. The exchange seeks to open a business bank account in a jurisdiction with recent digital currency regulations. The KYC analyst begins by querying \kgentity{CryptoX fraud} but finds no results. Realizing \kgentity{CryptoX} has a clean slate, the analyst shifts to peer-related checks using queries such as \kgentity{FTX} and \kgentity{Fraud}. This approach yields some results alongside \rollup options. Expanding the search to industry-wide topics like \kgentity{Bitcoin Exchange} and \kgentity{Financial Crime} produces a more comprehensive set of results, each linked to entities relevant to the chosen topics (highlighted in color). Armed with this information, the expert delves deeper into understanding the prevalent fraud types in the crypto realm and regulatory implications. Throughout this journey, the analyst enjoys the leeway to alternate between roll-up and drill-down modes, mirroring the flexibility of navigating an OLAP cube. In comparison, traditional due diligence would demand the painstaking manual tweaking of keywords and a thorough examination of search outputs to discern interconnected entities and patterns.

Our proposed OLAP operations for news exploration are versatile and hold potential for many other novel applications. For instance, \framework can detect media bias. When Elon Musk acquired Twitter, it stirred debates on wealthy individuals influencing media~\cite{twitter-musk}. By using \framework, users can expand from \kgentity{Elon Musk} to unveil parallels like Jeff Bezos's acquisition of the Washington Post~\cite{bezo-washington-post}, Patrick Soon-Shiong's purchase of the Los Angeles Times~\cite{patrick-la-times}, and Rupert Murdoch's buyout of The Wall Street Journal~\cite{murdoch-dow-jones}. While articles on Musk leaned negative, others retained a neutral or positive bent. These disparities underscore \framework's prowess in discerning biases and news narratives.

This OLAP-like interaction is enabled through the integration of the ontology network of Knowledge Graphs (KGs)~\cite{tanon2020yago,lehmann2015dbpedia}. As illustrated in Fig \ref{fig:kg_toy_sample}, the KG encompasses not only a comprehensive fact network of entities and relationships but also extensive ontology information. While vast ontology networks make \rollup and \drilldown operations possible, two major challenges arise. Firstly, in \rollup operations, identifying relevant news articles that correspond to selected roll-up concepts is complicated by the fact that roll-up concepts are part of the KG's ontology space, while the news articles' entities often belong to the KG's instance space.
To address this, we develop a relevance measure linking these two spaces and devised efficient path-based methods to quickly pinpoint relevant articles for roll-up analysis. Secondly, for \drilldown operations, the key challenge lies in the
effective ranking of sub-topic concepts to facilitate in-depth \drilldown investigations by analysts. This task is particularly demanding due to the vast array of potentially relevant concepts. To meet this challenge, we design a ranking framework according to the business objectives, incorporating three critical dimensions: coverage, specificity, and diversity. We summarize our technical contributions as follows:

\begin{itemize}[leftmargin=*]
    \item  We introduce a novel framework, \framework, which is the first of its kind, supporting news exploration with semantic \rollup and \drilldown operations by leveraging the ontology and fact networks of Knowledge Graphs (KGs).
    \item  We develop effective ranking schemes to evaluate concept-article relevance, enabling smooth \rollup and \drilldown operations. Additionally, we devise an efficient, unbiased sampling estimator for relevance score computation. 
    \item We assessed \framework using extensive news datasets, garnering feedback from Amazon Mechanical Turk's master-qualified participants. With over 3,900 evaluations, the findings affirm \framework's superiority over leading news search techniques. Even after integrating GPT models into all compared baselines, the results remained consistent.
    \item We also release our implementations, datasets, evaluation results and a full report at\cite{ncexplorersource}. Our dataset contains 200k news articles, with 2.9 million entity and 3.7 million concept annotations linked to DBPedia~\cite{dbpedia2016snapshot}. Compared to existing news datasets, the inclusion of entity and concept annotations enables deeper analysis of news articles, leveraging the interconnectedness of the knowledge graphs.
\end{itemize}

\section{Related Work}
\label{chap:related_work}

Our work is broadly related to news search and recommendation, aiming to enhance user experience in these domains. Existing works in this area typically extend generic natural language understanding models~\cite{devlin2018bert,reimers2019sentence,sheu2020context} to improve search and recommendation experiences for news articles. These works generally fall into two categories below: 

\noindent\textbf{Embedding-based approaches} employ dense vectors to generate document representations that encapsulate the semantics of news articles. Notable works include DKN~\cite{wang2018dkn}, KRED~\cite{liu2020kred}, and NewsGraph~\cite{newsgraph}. DKN integrates the embeddings of news titles, derived from a CNN approach~\cite{kim2018deep}, with the embeddings of news KG entities obtained through a knowledge graph embedding technique~\cite{transd}. KRED extends this idea further by introducing a context embedding layer generated from a news entity's neighbors, and an attentive merging layer inspired by the KG Attention Network (KGAT). NewsGraph builds a separate graph by removing irrelevant edges and introducing new edges that represent co-occurrences in the same news or visits from the same user. While embedding-based approaches are shown to improve the semantic representations of documents, they lack an explicit explanation to associate entities mentioned in the text with concepts appeared in the query, which is crucial for due diligence.

\noindent\textbf{Structure-based approaches:} Our work is more related to the methods that focus on utilizing the KG structures to provide better explainability and semantic relevance between news articles. AnchorKG~\cite{DBLP:conf/kdd/LiuLLWS021} and NewsLink~\cite{yang2021newslink} are two prominent works in this category. AnchorKG represents each news article as a compact subgraph containing essential news entities and their k-hop neighbors, generated using a reinforced learning framework. This interaction of anchor graphs provides explanations for semantic relevance between articles. NewsLink, a state-of-the-art news search approach, uses seed nodes identified from text fragments and a graph expansion algorithm to connect nodes in a single graph, adding hidden related nodes as auxiliary information for news semantic representation. \revision{These methods boost search performance via KG's fact network. However, their exploration capabilities are still limited due to lack of KG ontology integration.}
To the best of our knowledge, our work is the first to facilitate systematic exploration with OLAP-like operations. Furthermore, there are several works on using general knowledge bases to generate outstanding facts~\cite{fan2017discovering,yang2021context}. These approaches can be integrated into our system with ease.


\section{\framework}
\label{chap:architecture}

\begin{figure*}
    \vspace{-0.5cm}
    \centering
    \begin{minipage}{.4\linewidth}
        \centering
 \includegraphics[width=\textwidth]{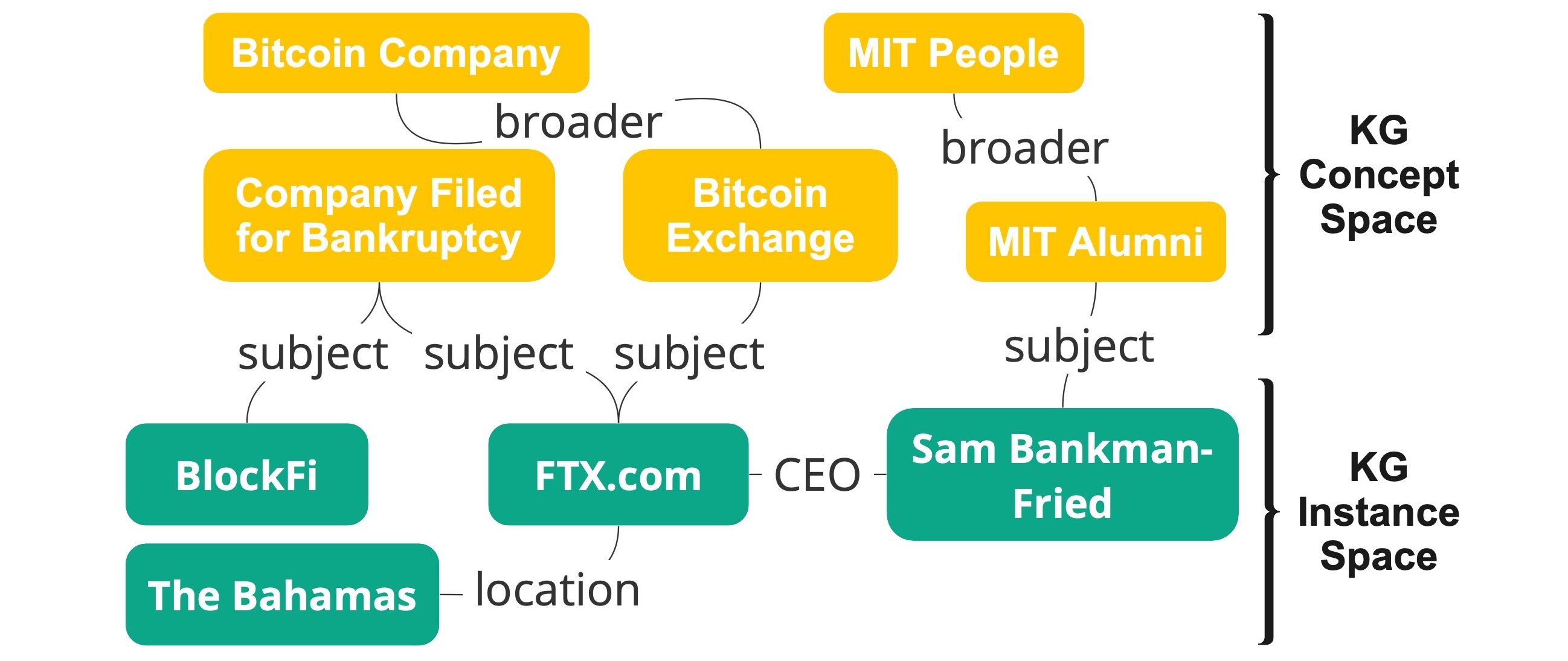}
    \caption{KG Concept and Instance Spaces}
    \label{fig:kg_toy_sample}
    \end{minipage}
    \hspace{1mm}%
    \begin{minipage}{.55\linewidth}
        \centering
       \includegraphics[width=\textwidth]{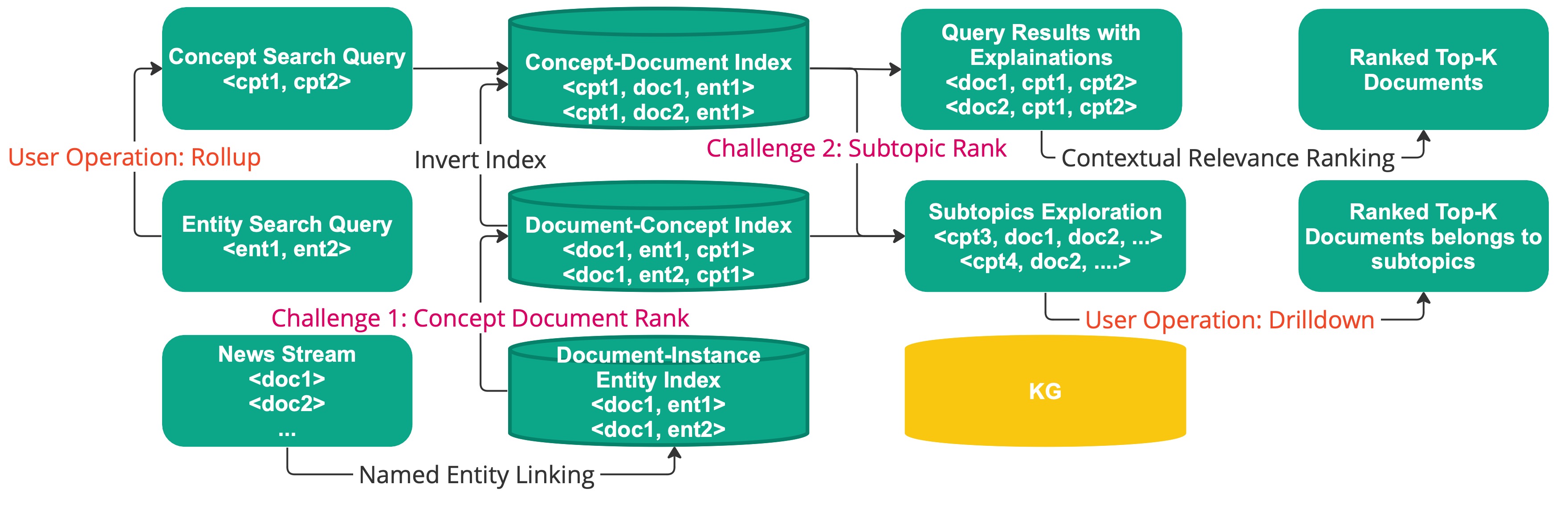}
   \caption{\framework Architecture.}
   \label{fig:architecture}
    \end{minipage}
 \vspace{-2em}
\end{figure*}

\noindent \framework leverages external KGs for news exploration.
A KG is a multigraph $\mathcal{G}=(\vsetcon \bigcup \vsetins,\esetcon \bigcup \esetins,\onto)$. $\vsetcon$ and $\vsetins$  represent the concept entities and the instance entities, represented by yellow and green nodes in Fig~\ref{fig:kg_toy_sample} respectively. Each concept edge in $\esetcon$ links two concept entities, and an instance edge in $\esetins$ links two instance entities. Like NewsLink~\cite{yang2021newslink}, we add a reversed edge for each original edge so that $\mathcal{G}$ is bidirected. The associations between the instance space and the concept space are captured by the ontology relation $\onto$. $\onto(c)$ maps a concept entity $c \in \vsetcon$ to a set of instance entities while $\onto^{-1}(v)$ maps an instance entity $v \in \vsetins$ to a set of concept entities. \revision{Fig.~\ref{fig:architecture} shows the framework's architecture. When a stream of news articles arrives, they undergo  a pipeline of natural language processors (NLP) of tokenization, entity recognition and entity linking. We use the spaCy~\cite{honnibal2017spacy} library as our NLP tool to transform a document into a list of KG instance entities. The next step is to link a document with the KG ontology space to enable two major components of \framework: semantic \rollup and \drilldown.}

\subsection{Roll-up Operation.}
\label{sec:ranking}
\noindent \framework generates entities from documents $d$, allows users to replace some entities with KG \revision{concepts, and further \rollup along the edge type \kgentity{broader}}, to form a concept pattern query $Q$. Given such a $Q$, a document $d$ matches $Q$ if for each concept $c \in Q$, there is an entity $v$ in $d$ that $v \in \onto(c)$. In other words, we can find a set of entities in $d$ that match $Q$. 
The relevance score of a matched document $d$ to $Q$ is the sum of the relevance scores among all concepts in $Q$, i.e.,  
\begin{equation}
	rel(Q,d) = \sum_{c \in Q} cdr(c,d).
\end{equation}

\revision{
\begin{definition}\label{def:rollup}[Roll-up]
Given a concept pattern query $Q$, return top-K documents $d$ with the highest score $rel(Q, d)$.
\end{definition}
}


\noindent \textbf{Concept Document Rank}.~We propose a novel ranking scheme for $cdr(c,d)$ by considering two key relevance dimensions: \emph{ontology relevance} $cdr_o(c,d)$ and \emph{context relevance} $cdr_c(c,d)$.
A concept $c$ is relevant to a document $d$ if $c$ is relevant in both dimensions:
\vspace{-2mm}
\begin{equation}\label{eq:cdr}
	cdr(c,d) = cdr_o(c,d) \cdot cdr_c(c,d).
\end{equation}
\vspace{-4mm}

\subsubsection{Ontology Relevance}
When a document $d$ contains an entity $v$ that matches a concept $c$ under the ontology relation $\onto$, i.e., $v \in \onto(c)$, $c$ is associated with $d$ by ontology relevance. Since there could be more than one entity in $d$ that matches $c$, 
we define the ontology relevance score function as follows:
\begin{equation}\label{eq:cdr_o}
	cdr_o(c,d) = log\frac{|\vsetins|}{|\onto(c)|} \cdot \left(\max_{v \in ME(c,d)} tw(v,d)\right)
\end{equation}
where $ME(c,d) =  \{ v | \; v \in d \;\text{and}\; v \in \onto(c) \}$.
First, a concept that matches more entities in the ontology, i.e., lower specificity, should be less relevant to a document. Second, among the matched entities, a \emph{pivot} entity is selected as the one with the highest term weight $tw(v,d)$ in the document to match the concept. The term weight reflects  the importance of $v$ in $d$. If $v$ plays a more significant role in $d$, $c$ is more relevant to $d$. We use the typical TF-IDF scheme for term weighting in implementation and other schemes can be easily supported. \revision{For a broad concept that does not have a direct link to document entities, $cdr_o(c, d)$ is replaced with an edge concept among its children that matches document entities}.

\subsubsection{Context Relevance}

An entity in a document could map to different concepts by ontology. 
Hence, ontology relevance can only differentiate the matched concepts of the same entity with the specificity score $\log(|\vsetins|/|\onto(c)|)$, which simply penalizes broad concepts.
We thus propose to include the unmatched entities as contextual information to improve the relevance semantics. 
To measure the context relevance of $c$ and $d$, we compute the relevance of $c$ and the context entities $CE(c,d) =  \{ v | \; v \in d \;\text{and}\; v \notin \onto(c) \}$. 
Although the context entities do not match $c$, we take advantage of both the ontology relation and the instance space to link $c$ with the context entities.
We introduce a novel connectivity score $conn(c,d)$ to measure the KG connectivity between a concept $c$ and a context entity set $CE(c,d)$ as follows:
\vspace{-2mm}
\begin{equation}\label{eq:conn}
	conn(c,d) =  \sum_{v \in CE(c,d)} 
	\frac{\sum_{u \in \onto(c)}\sum_{l=1}^\tau \beta^l \cdot |paths_{u, v}^{<l>}|}{|CE(c,d)|}
\end{equation}
where $|paths_{u, v}^{<l>}|$ measures the number of $l$-hop simple paths connecting $u$ and $v$ in the instance space, and $\beta$ is a damping factor that penalizes longer paths. 
Intuitively, the connectivity score is the average number of paths among all context entities connected to any KG instance entity that matches $c$, subject to a hop constraint of at most $\tau$. Thus, better connectivity leads to a higher relevance score.
Finally, we normalize the connectivity score to $[0,1)$ as follows:
\vspace{-2mm}
\begin{equation}\label{eq:cdr_c}
	cdr_c(c,d) = 1 - \frac{1}{1+conn(c,d)}
\end{equation}

\subsection{Drill-down Operation.}
\label{sec:subtopic}
\noindent Based on the matched news from the \rollup operation,
\framework suggests additional concepts as subtopics to a query $Q$. 
A suggested concept $c'$, as a subtopic of $Q$, enables users to conduct a \emph{drill-down} analysis.
By selecting $c'$, users narrow down the matched news to the augmented query $Q \cup c'$. Since a subtopic must appear as a concept in at least one matched document of $Q$, we can find all candidate subtopics by unionizing the concepts from all matched documents.

Let $\mathcal{D}(Q)$ denote the set of retrieved documents for $Q$ and a candidate subtopic is a concept $c$ such that there is an instance entity $v \in \onto(c)$ appearing in one of the documents $d \in \mathcal{D}(Q)$.
We develop a score $sbr(\cdot,Q)$ over candidate subtopics and only suggest top-scored subtopics for $Q$. 
There are three key perspectives considered in $sbr(\cdot,Q)$ as below:
\begin{equation*}
	\fontsize{9}{9}sbr(c,Q)=coverage(c,Q) \cdot specificity(c) \cdot diversity(c,Q)
\end{equation*}


    \subsubsection{Coverage} 
    The coverage of a subtopic $c$ is calculated as the sum of the relevance scores $cdr(c, d)$ for all documents $d$ in $\mathcal{D}(Q)$. This ensures that only subtopics that are highly relevant to a large number of documents are suggested to the user for further exploration: $coverage(c,Q)=\sum_{d \in \mathcal{D}(Q)} cdr(c,d)$.
 \subsubsection{Specificity} To avoid suggesting trivial subtopics like \kgentity{Person} which may match a large number of entities, we prioritize concepts with higher specificity scores using the formula $specificity(c)=\log(|\vsetins|/|\onto(c)|)$. 

 \subsubsection{Diversity} The diversity score calculates the average number of distinct entities mapped to the concept $c$ among all the documents relevant to $Q \cup \{c\}$: $diversity(c,Q) = \frac{|\bigcup_{d \in \mathcal{D}(Q)} ME(c,d)|
		}{|\mathcal{D}(Q \cup \{c\})|}$.
 This helps ensure fairness in the suggested concepts and prevents results from being biased toward concepts that match a small set of popular entities.
\revision{
\begin{definition}\label{def:drilldown}[Drill-down]
Given a concept pattern query $Q$, return top-K subtopics $c$ with the highest $sbr(c,Q)$ score.
\end{definition}
}

\subsection{Connectivity Score Estimation}
\label{sec:connectivity_score_estimation}
The connectivity score can be computationally expensive due to the massive number of path enumerations for each pair of entities and existing studies on s-t path enumeration can only produce \emph{polynomial delay} algorithms~\cite{qin2019towards,sun2021pathenum}. Inspired by the sampling approaches for online join aggregation queries~\cite{li2016wander,zhao2018random,park2020g,sun2021thunderrw}, we devise 
a single random walk estimator. 
To start a random walk, we first sample a source entity $u$ from all entities that map to the concept $c$, and a target entity $v$ from the corresponding context entities.
A \emph{non-repeating} random walk $r(u,v)$ from $u$ tries to reach $v$ samples via at most $\tau$ distinct nodes. 
Let $u_i$ denote the $i$th node sampled by a random walk starting from the source $u$ and $N(u_i)$ is the number of eligible neighbors of $u_i$ to be sampled, we define the following sample estimator:
 \vspace{-1em}
\begin{equation}
	r(u,v) = \mathcal{I}(u,v) \cdot |\onto(c)| \cdot \beta^{l-1} \cdot \displaystyle \prod_{i=1}^{l-1} N(u_i)
\end{equation}
     \vspace{-1em}

where $\mathcal{I}(u,v)$ is an indicator random walk: it returns $1$ if $u$ reaches $v$ at the $l$-th sampled node for any $l \leq \tau$; otherwise, it returns $0$. The sampling process may suffer from slow convergence, especially when many of the sampled paths cannot reach the target context entity $v$.  
To enable faster convergence, we build a reachability index~\cite{cheng2014efficient} on the KG instance space and only sample eligible neighbors that satisfy the hop constraint. 
\if\fullversion1
The following theorem ensures our sample approach is unbiased.
\begin{theorem}
	$r(\cdot,\cdot)$ is an unbiased estimator to $conn(c,d)$, i.e., $\mathbb{E}[r(\cdot,\cdot)] = conn(c,d)$ where the randomness is taken over the source $u$ and target $v$ as well as the random walk from $u$ to $v$. 
\end{theorem}
Proof Sketch. $conn(c,d)$ is a weighted sum of all the paths connecting an entity node $u \in \onto(c)$ and a context entity $v \in CE(c,d)$ subject to a hop constraint of $\tau$. For each path $s=\{u_1=u,u_2,u_3,...,u_l=v\}$ that connects $u$ and $v$, the probability of $s$  being sampled is $\mathbb{P}(s)=(|\onto(c)|\cdot|CE(c,d)|\cdot\prod_{i}^{l-1} N(s_i))^{-1}$.
Thus, we can use the Horvitz–Thompson estimator~\cite{HTestimator} to approximate the population sum without bias.

\begin{algorithm2e}[t]
	\SetKwInOut{Input}{Input}
	\SetKwInOut{Output}{Output}
	\Input{KG $\mathcal{G}$, concept $c$, document $d$, sample size $\theta$}
	\Output{the estimated connectivity $est$}
        $est \gets 0$; \\
	\While{there are less than $\theta$ random walks}{
		  $l \gets 1$; $p \gets 1.0$; \\
		  $u_l \gets$ a random entity in $\onto(c)$; \\
		  $v \gets$ a random entity in $CE(c,d)$; \\
		  \While{$l \leq \tau$}{
			$count \gets 0$; \\ 
			\ForEach{$n \in N(u_l)$}{
				\If{$hop(n,v) \leq \tau-l$ \label{alg:hop-constrain}}{
					$count \gets count + 1$; \\
					$u_{l+1} \gets n$ with probability $\frac{1}{count}$;\label{alg:single-scan}
				}				
			}
			$p \gets \frac{p}{count}$; \\
			$l \gets l+1$; \\
			\textbf{break} on $u_{l} = v$; 
		}
		\If{$u_{l} = v$}{
			$est \gets est + \frac{\beta^{l-1}}{p}$;
		}
	}
	\Return {$\frac{est \cdot |\onto(c)|}{\theta} $.} \\
	\caption{Random walk estimator with k-hop index.} 
	\label{algo:samplekhop}
\end{algorithm2e}

\vspace{1mm}
\noindent\textbf{Convergence Optimization}.~Although we devise an unbiased estimator for the connectivity score, the sampling process may suffer from slow convergence, especially when many of the sampled paths cannot reach the target context entity $v$.  
To enable faster convergence, we build a reachability index~\cite{cheng2014efficient} on the KG instance space and only sample eligible neighbors that satisfy the hop constraint. We present our sampling approach with the k-hop reachability index in Algorithm~\ref{algo:samplekhop}.
For each random walk, we first sample the source $u_1$ and the target $v$. 
Subsequently, we iteratively sample the next node $u_{l+1}$ by scanning the neighbors of $u_l$. With the help of the k-hop index, we first check if a neighbor $n$ can reach $v$ within the hop constraint of $\tau-l$ (Line \ref{alg:hop-constrain})
Then, we uniformly sample eligible neighbors by a Reservoir sampler, which only requires a single scan of the neighbors (Line \ref{alg:single-scan})
A random walk terminates when either the target $v$ is sampled or the hop limit is reached. 
\else
\revision{The theorem and proof of our sample approach is unbiased can be found in the full report \cite{ncexplorerfull}}.
\fi

\section{Evaluation}
\label{chap:evaluation}
\if\fullversion1
We present the  setup in Sec.~\ref{sec:exp-setup} and then answer the following questions: 
\begin{enumerate}
    \item Can \framework produce relevant results given topics rolled up from news articles? (Sec.~\ref{sec:relevance-study})
    \item Can \framework efficiently process large-scale news corpus and knowledge graphs? (Sec.~ \ref{sec:performance-study})
    \item Can \emph{context relevance score} effectively measure the relevance between a concept and a document? How does the proposed sampling method impact the accuracy? (Sec.~ \ref{sec:csc-effectiveness-study})
    \item How effective is each scoring component in the ranking model for the \drilldown operation? (Sec.~\ref{sec:subtopic-ranking-study})
    \item How the combination of \rollup and \drilldown operations improve the productivity of due diligence and risk management tasks in financial institutions.(Sec.\ref{sec:rollup-drilldown-study})
    \item Can \framework support real analytical applications? (Sec.~\ref{sec:case-study})
\end{enumerate}
\fi

\if\fullversion1
\subsection{Settings}\label{sec:exp-setup}
\fi
\noindent\textbf{Datasets.}
We use the June 2021 snapshot of DBPedia~\cite{dbpedia2016snapshot} as our backend KG. We crawl 200k articles from popular news portals: \emph{Reuters}~\cite{reuters}, \textit{SeekingAlpha}~\cite{seekingalpha} and \textit{The New York Times}~\cite{nytimes} to have a mixture of business and politics reports. 
Detailed statistics of released dataset are shown below.
\begin{table}[ht]
    \vspace{-2mm}
    \centering
    \begin{adjustbox}{width=1\linewidth}
        \begin{tabular}{ |c| c c c|}
        \hline
         News Source & Articles & Total Entities & Linked Entities \\ \hline
         Seekingalpha & 6823 & 97k   & 62k (63.9\%)\\
         NYT & 3625 & 51k & 35k (68.6\%) \\  
         Reuters & 171662 & 4539k & 2336k (51\%) \\
         \hline
        \end{tabular}
    \end{adjustbox}
    \vspace{-2mm}
\end{table}

\noindent\textbf{Compared Methods.}
\begin{itemize}[leftmargin=5mm]
	\item \lucene implements a typical bag-of-words keyword match model. We use BM25~\cite{robertson2009probabilistic} for the term weighting scheme with the default library settings.
	\item \bert~\cite{devlin2018bert} is a popular neural text embedding model. We use SBERT~\cite{sbert-model}, a modification of the pre-trained \bert to map each news article to a vector of 768 dimensions.
	\item \newslink~\cite{yang2021newslink} is the state-of-the-art implicit news exploration method that expands a news document and a query by forming a common ancestor graph extracted from the KG. Each KG entity in the extracted graph is then treated as a matching keyword in the bag-of-words model. 
	\item \newsbert is a hybrid method that combines \newslink and \bert. It expands query entities into a subgraph using \newslink's algorithm and concatenates them to form a long text query. 
	\item \framework is our proposed approach. The parameters are set to $\tau=2$ and $\beta=0.5$ by default. The number of samples for connectivity score estimation is set to 50. 
\end{itemize}

\vspace{1mm}\noindent\textbf{Implementation.}
\framework and \newslink are implemented in Python 3.9. 
\bert and \newsbert use Qdrant~\cite{qdrant} as vector search engine. We use a server running on Ubuntu 20.04 with an AMD EPYC 7643 Processor @ 3.45GHz and 251G RAM. Our sourcecode is released at\cite{ncexplorersource}.

\begin{table*}[ht]
\centering
  \begin{minipage}{0.70\linewidth}\centering
\caption{NCDG@K without GPT represents each method's original top-k results. NCDG@K with GPT represents the performance after GPT re-ranks the top-k results. The best results are boldfaced and the second-best results are underlined.}

\begin{adjustbox}{width=1\textwidth}
\begin{tabular}{ |c|P{1.6cm}P{1.6cm}P{1.6cm}|P{1.6cm}P{1.6cm}P{1.6cm}|} 
    \hline
 & NDCG@1 & NDCG@5 & NDCG@10 & NDCG@1 & NDCG@5 & NDCG@10 \\
  \hline
 & wo/w the GPT rerank & wo/w the GPT rerank & wo/w the GPT rerank  & wo/w the GPT rerank & wo/w the GPT rerank & wo/w the GPT rerank \\
    \hline
    Topic & \multicolumn{3}{c|}{\textbf{International Trade}} & \multicolumn{3}{c|}{\textbf{Lawsuits}}  \\ 
        \hline
    
    Lucene &  0.688 / 0.572 & 0.557 / 0.532 & 0.737 / 0.720 & 0.574 / 0.571 & 0.593 / 0.597 & 0.763 / 0.766\\
    BERT &   \underline{0.856} / \underline{0.856} & \underline{0.882} / \underline{0.882} & \underline{0.951} / \underline{0.951} & \textbf{0.849} / \underline{0.849} & \underline{0.848} / \underline{0.900} & \underline{0.935} / \underline{0.949} \\
    NewsLink &  0.765 / 0.817 & 0.623 / 0.650 & 0.781 / 0.799 & 0.329 / 0.628 & 0.406 / 0.475 & 0.636 / 0.683 \\
    NewsLink-BERT & 0.825 / 0.836 & 0.877 / 0.878 & 0.949 / 0.949 & 0.627 / 0.658 & 0.796 / 0.848 & 0.883 / 0.898 \\
    \framework &  \textbf{0.974} / \textbf{0.974} & \textbf{0.957} / \textbf{0.956} & \textbf{0.987} / \textbf{0.986} & \underline{0.844} / \textbf{1.000} & \textbf{0.919} / \textbf{0.948} & \textbf{0.959} / \textbf{0.979} \\
    \hline

    Topic & \multicolumn{3}{c|}{\textbf{Elections}} & \multicolumn{3}{c|}{\textbf{Mergers \& Acquisitions}} \\ 
  
    \hline
    Lucene & 0.550 / 0.273 & 0.455 / 0.378 & 0.653 / 0.603 & 0.464 / 0.659 & 0.593 / 0.629 & 0.777 / 0.802  \\
    BERT & 0.887 / 0.910 & 0.894 / 0.903 & 0.941 / 0.947 & \textbf{0.728} / \textbf{1.000} & \underline{0.820} / \textbf{0.879} & \underline{0.915} / \underline{0.955} \\
    NewsLink & 0.554 / 0.554 & 0.450 / 0.466 & 0.649 / 0.660 & 0.305 / 0.493 & 0.445 / 0.471 & 0.675 / 0.693 \\
    NewsLink-BERT & \textbf{0.946} / \underline{0.946} & \textbf{0.972} / \textbf{0.972} & \textbf{0.990} / \textbf{0.990} & \underline{0.724} / \underline{0.912} & 0.803 / 0.829 & 0.913 / 0.930 \\
    \framework &  \underline{0.924} / \textbf{0.947} & \underline{0.958} / \underline{0.966} & \underline{0.978} / \underline{0.984} & 0.712 / 0.846 & \textbf{0.843} / \underline{0.870} & \textbf{0.937} / \textbf{0.956}\\
    \hline
    Topic & \multicolumn{3}{c|}{\textbf{International Relations}} & \multicolumn{3}{c|}{\textbf{Labor Dispute}} \\ 
    \hline
    Lucene &   0.896 / 0.650 & 0.722 / 0.670 & 0.830 / 0.795 & 0.564 / 0.621 & 0.618 / 0.632 & 0.817 / 0.826   \\
    BERT &  \underline{0.921} / \underline{0.957} & 0.922 / 0.941 & 0.959 / 0.970 & 0.370 / 0.370 & 0.411 / 0.410 & 0.670 / 0.669 \\
    NewsLink & 0.735 / 0.804 & 0.729 / 0.760 & 0.834 / 0.854 & 0.481 / 0.729 & 0.476 / 0.499 & 0.716 / 0.733 \\
    NewsLink-BERT & 0.867 / 0.945 & \underline{0.943} / \underline{0.954} & \underline{0.971} / \underline{0.982} & \underline{0.695} / \underline{0.905} & \underline{0.720} / \underline{0.767} & \underline{0.889} / \underline{0.922} \\
    \framework & \textbf{0.927} / \textbf{0.963} & \textbf{0.970} / \textbf{0.974} & \textbf{0.986} / \textbf{0.989} & \textbf{0.922} / \textbf{0.931} & \textbf{0.984} / \textbf{0.988} & \textbf{0.989} / \textbf{0.991}  
 \\
    \hline
\end{tabular}
\end{adjustbox} 
  \label{tab:similarity-evaluation-result}
\end{minipage}
\begin{minipage}{0.29\linewidth}
  \caption{Impact of the GPT-rerank. }
\begin{adjustbox}{width=1\linewidth}

\begin{tabular}{ |P{1.8cm}|R{0.97cm}R{0.82cm}R{0.92cm}|} 
    \hline
  & \scriptsize{NDCG@1} & \scriptsize{NDCG@5} & \scriptsize{NDCG@10}\\
 \hline
 Lucene & -10.44\% & -2.77\% & -1.40\%  \\
 BERT & 7.18\% & 2.89\% & 1.32\% \\
 NewsLink & 27.01\% & 6.20\% & 3.08\%  \\
 \scriptsize{NewsLink-BERT} & 7.05\% & 2.17\% & 1.05\% \\
 \scriptsize{NCEXPLORER} & 6.75\% & 1.26\% & 0.84\% \\
 \hline
\end{tabular}
\end{adjustbox} 
  \label{tab:gpt-ndcg-impact}
  \vspace{-2.6mm}
\caption{
  \rollup and \drilldown effectiveness study. Number of results generated within the same time limit. 
}
  
 \begin{tabular}{ |P{0.3cm}|P{1.2cm}|P{1.2cm}|P{0.9cm}| } 
    
    \hline
    
     Task No. & Keyword Search (avg/std) & NC- explorer (avg/std)  & p-value of H1 (n=10) \\
    \hline
     1 & 0.6/1.07 & \textbf{2.7}/1.56 & 0.016\\
     2 & 0.5/0.97 & \textbf{4.0}/1.41 & 0.004\\
     3 & 0.9/0.99 & \textbf{2.8}/0.46 & 0.001\\
     4 & 0.9/1.10 & \textbf{2.7}/0.67 & 0.001\\
     5 & 0.7/0.95 & \textbf{1.8}/0.42 & 0.007 \\
     6 & 1.0/1.56 & \textbf{4.5}/1.65 & 0.002 \\
     7 & 1.0/1.41 & \textbf{4.2}/1.03 & 0.001\\
     8 & 1.8/1.03 & \textbf{3.3}/0.95 & 0.002\\

  \hline

\end{tabular}
  \label{tab:productivity-gain-study}
\end{minipage}
    \vspace{-5mm}
\end{table*}

\subsection{Roll-up Operation}
\subsubsection{Concept Document Rank Evaluation}\label{sec:relevance-study}
As shown in Table~\ref{tab:similarity-evaluation-result}, we evaluate a total of six topics. Each topic is combined with either an entity group (a list of countries or companies) to form queries such as \kgentity{Elections in African countries} or \kgentity{Lawsuits involving U.S. technology companies}. For each query, the top 5 news articles are retrieved from each method, resulting in 25 outcomes. These results are presented to each evaluator in a randomized order. To ensure a fair comparison, we hide \framework's result explanations. The relevance level is rated for each concept in the query, with values ranging between 0 and 5. In total, we obtain 3,900 ratings from 78 evaluators. We use \textbf{NDCG@K} to evaluate the effectiveness of relevance ranking. The GPT models~\cite{gpt-3.5-turbo, gpt-4} offer unparalleled text analysis capabilities. \revision{To explore whether we can use GPTs to further boost the performance, we input results from each method into GPT 3.5-turbo~\cite{gpt-3.5-turbo} to re-rank the relevance between a topic and a news article based on the prompt:}

\noindent\fbox{
  \parbox{0.95\linewidth}{%
    $<$news article$>$. Is this article related to $<$topic$>$. please give a rating between 0.000 and 5.000, with 5.000 being most relevant. only give three decimal digits.
    }
}

\noindent The new \textbf{NDCG@K} scores are displayed on the right-hand side. The result for both scenarios shows that \framework achieves the best or second-best performance in nearly all cases, except for the query \kgentity{Merger \& Acquisition, U.S. biotechnology companies}. A detailed analysis reveals that evaluators show greater confidence in commonly known surface words like \kgentity{M\&A}, \kgentity{acquisition}, \kgentity{buy}, and \kgentity{sell}, while expressing uncertainty about specialized terms such as \kgentity{takeover}. This finding highlights the effectiveness of \framework's roll-up operation in helping analysts collect news articles featuring more domain-specific vocabularies. Embedding models also demonstrate remarkable performance in news exploration tasks, particularly when combined with KG information. However, there are two issues that arise when using pure embedding models. First, implicit matching may retrieve reports like daily trade price/volume\cite{bloomberg-stock-futures}, which, although crucial for real-time decision-making, have limited utility in exploration tasks. Second, the issue emerges when the concept entity isn't frequently mentioned alongside instance entities. For example, \kgentity{labor dispute} is often reported in the news as strikes organized by various labor unions. The embedding of \kgentity{labor dispute} alone cannot yield high-quality results. This issue can be partially mitigated with supplementary context provided by Newslink's subgraph KG embedding. For \newslink, the performance is not stable due to the infrequent formation of densely connected single components by the subgraph embedding of multiple concept entities. Instead of identifying hidden nodes that connect existing query entities, the subgraph often results in a single concept entity accompanied by its N-hop neighbors. This dilutes the significance of concepts with smaller connected components, affecting the overall effectiveness of the news exploration process. Table~\ref{tab:gpt-ndcg-impact} shows the impact of incorporating the GPT model to re-rank retrieved documents. One observation is that the impact is positive for all methods except \lucene. Another interesting observation is the impact for NDCG@1 $>$ NDCG@5 $>$ NDCG@10, suggesting that GPT can distinguish the subtle differences among top results. \revision{Whether it is feasible to use GPT directly as a relevance ranker instead of a re-ranker of retrieved results is a topic for our upcoming research.}

\if\fullversion1
\noindent \textbf{Rating Scale Impact Study} Existing work has shown that the effect of scale used in relevance ranking study for crowdsourced participants do not affect final results\cite{effect_of_relevance_scales}. To validate this finding for GPT reranker. We repeated above GPT experiment with 0-10 scale. An independent samples t-test against original 0-5 scale experiment shows a p-value of 0.878. Our chosen significance
level (alpha) is 0.05. The p-value is bigger than alpha, rejecting the null hypothesis, showing different scales do not affect evaluation results.
\fi

\subsubsection{Efficiency Study}\label{sec:performance-study}

\if\fullversion0
\begin{figure*}
 \begin{adjustwidth}{-0.5cm}{-0.5cm}

    \centering
    \begin{minipage}{.18\linewidth}
        \centering
        \includegraphics[width=\textwidth]{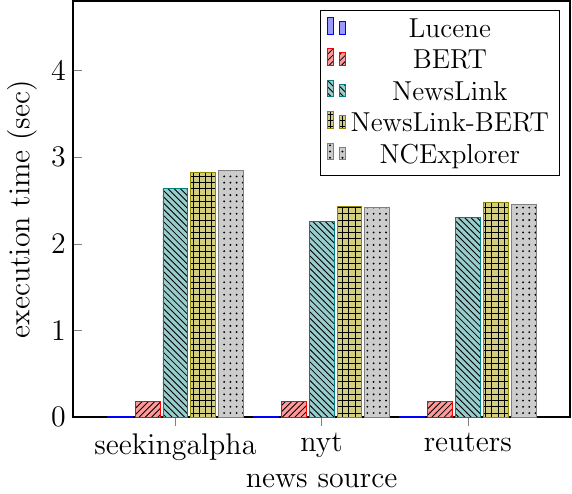}
        \caption{Performance Study: indexing time}
        \label{fig:document_indexing_time}
    \end{minipage}
    \hspace{1mm}%
    \begin{minipage}{.18\linewidth}
        \centering
        \includegraphics[width=\textwidth]{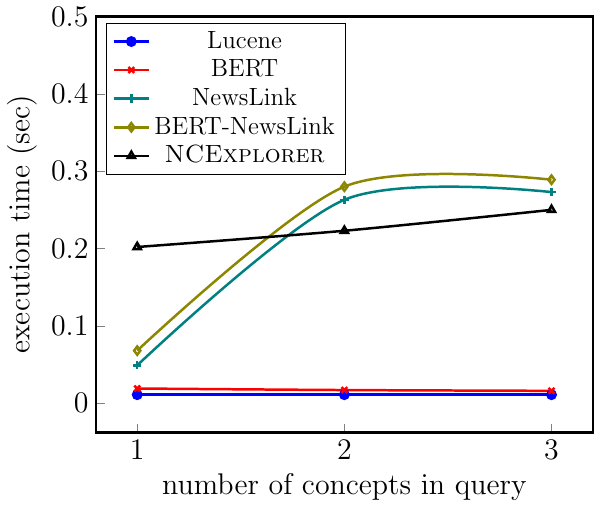}
          \caption{Performance Study: \\retrieval time}
            \label{fig:document_retrieval_time}
    \end{minipage}
    \hspace{1mm}
    \begin{minipage}{.18\linewidth}
        \centering
        \includegraphics[width=\textwidth, trim={0 0 0 0},clip]{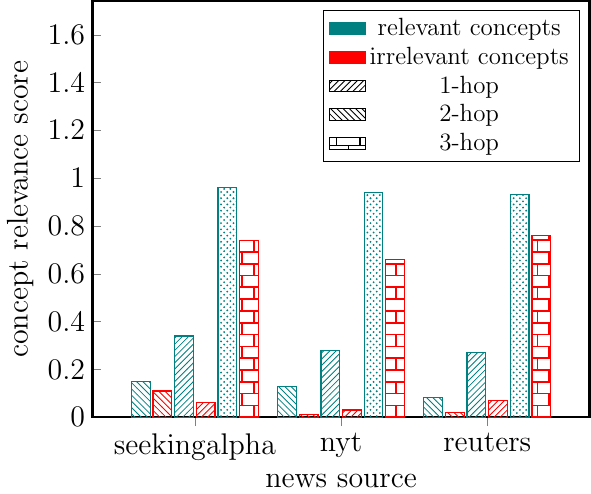}
        \caption{Effectiveness study: context relevance}
        \label{fig:csc-effectiveness}
    \end{minipage}
    \hspace{1mm}%
    \begin{minipage}{.18\linewidth}
        \centering
        \includegraphics[width=\textwidth, trim={0 0 0 0},clip]{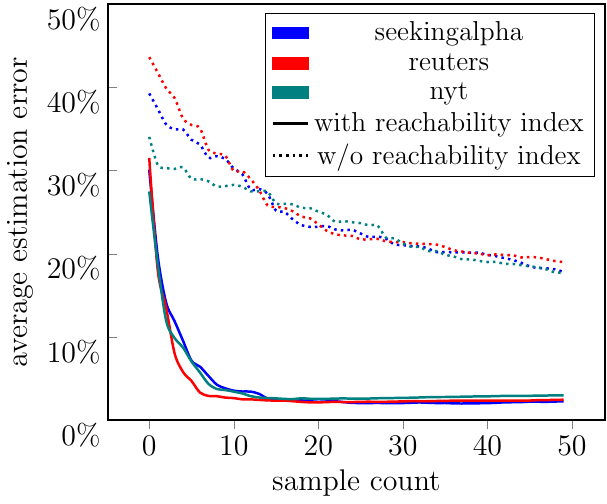}
        \caption{Sampling error with increasing samples}  
        \label{fig:csc-sampling-error}
    \end{minipage}
    \hspace{1mm}
    \begin{minipage}{.18\linewidth}
        \centering
        \includegraphics[width=\textwidth]{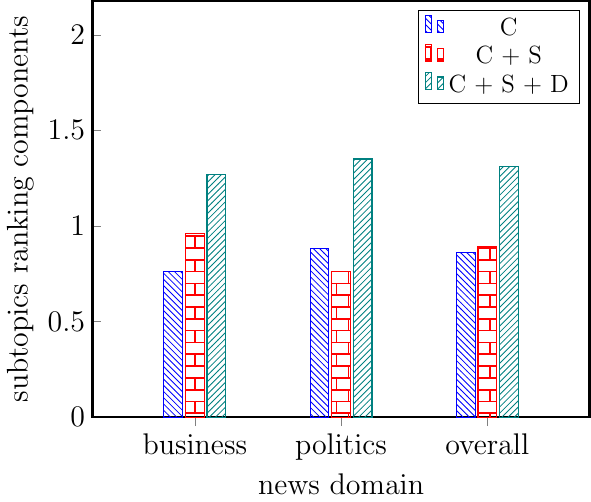}

        \caption{Subtopics ablation study}
        \label{tab:subtopic_ranking_survey_result}
    \end{minipage}
 \end{adjustwidth}
\vspace{-6mm}
\end{figure*}

\else
\begin{figure*}[!htb]
    \begin{minipage}{.24\linewidth}
        \centering
        \includegraphics[width=\textwidth]{images/output-figure0.pdf}
        \caption{Indexing time}
        \label{fig:document_indexing_time}
    \end{minipage}
    \hspace{1mm}%
     \begin{minipage}{.4\linewidth}
        \centering
        \includegraphics[width=\textwidth]{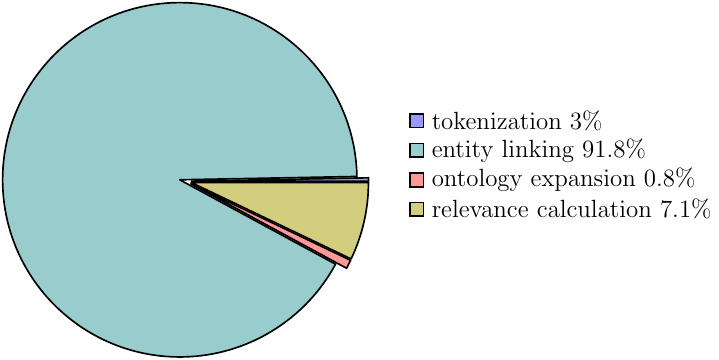}
        \caption{Document processing time}
        \label{fig:document_processing_time}
    \end{minipage}
    \begin{minipage}{.24\linewidth}
        \centering
        \includegraphics[width=\textwidth, trim={0 0 0 0},clip]{images/output-figure5.pdf}
        \caption{Sampling error study}  
        \label{fig:csc-sampling-error}
    \end{minipage}
    \begin{minipage}{.24\linewidth}
        \centering
        \includegraphics[width=\textwidth]{images/output-figure2.pdf}
          \caption{Retrieval time}
            \label{fig:document_retrieval_time}
    \end{minipage}
     \hspace{1mm}%
    \begin{minipage}{.24\linewidth}
        \centering
        \includegraphics[width=\textwidth]{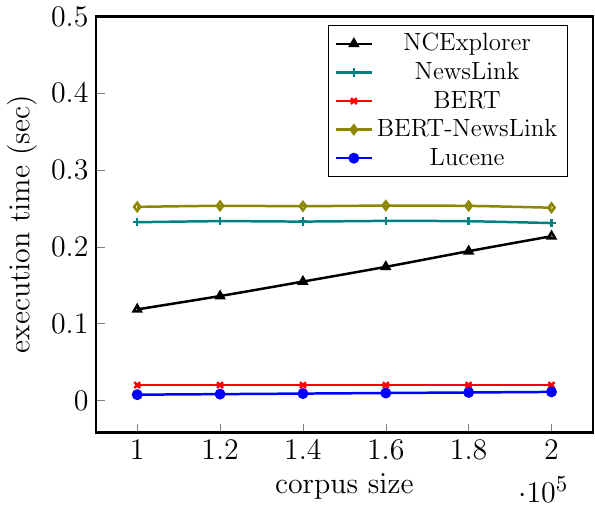}
        \caption{Scalability Study}
            \label{fig:scability-study}
    \end{minipage}
       \hspace{1mm}
      \begin{minipage}{.24\linewidth}
        \centering
        \includegraphics[width=\textwidth, trim={0 0 0 0},clip]{images/output-figure4.pdf}
        \caption{Context relevance study}
        \label{fig:csc-effectiveness}
    \end{minipage}
    \hspace{1mm}
    \begin{minipage}{.24\linewidth}
        \centering
        \includegraphics[width=\textwidth]{images/output-figure7.pdf}

        \caption{Subtopics ablation study}
        \label{tab:subtopic_ranking_survey_result}
    \end{minipage}
\end{figure*}

\fi

\noindent\textbf{Indexing Efficiency.} \framework processes each news document and constructs an index for query processing. To evaluate the indexing overhead, we select 100 articles from each news portal and report the average processing time in Fig.~\ref{fig:document_indexing_time}. 
\lucene and \bert show sub-second execution time.
\newslink, \newsbert, and \framework cost 2-3 seconds for each article. A breakdown analysis of the total cost showed that the top two overheads come from entity linking (91.8\%) and relevance score calculation (7.1\%). Entity linking is a common cost to all methods that require KG analysis. 
The primary focus of this work is not to improve the efficiency of entity linking. In fact, the calculation of relevance scores, which constitutes 7.1\% of the total indexing time, benefits from our efficient estimation approach using sampling techniques.
\revision{For constructing the reachability index on the DBpedia KG, comprising 5.2 million nodes and 27.9 million edges, the process takes 260 seconds and necessitates 100GB of memory.}

\vspace{1mm}
\noindent\textbf{Retrieval Efficiency.}
Fig~\ref{fig:document_retrieval_time} displays the query efficiency of the compared methods when increasing the number of concepts while keeping corpus size fixed. The processing time for each data point is the average across 100 queries.
\lucene is a highly optimized system for text retrieval and takes the least amount of time. \bert used to take longer execution time due to the absence of index. Recent development on vector databases has greatly sped up embedding retrieval and result in \lucene compatible speed.
The performance for \framework is similar to \newslink where the duration correlates to number of KG entities in the query. \newsbert takes the sum of \bert and \newslink. Overall, \framework can answer a concept pattern query with reasonable overhead. 
\if\fullversion1
Fig.~\ref{fig:scability-study} shows the relationship between retrieval time and corpus size for the set of queries. \lucene and \bert achieve the best scalability thanks to open source solutions. \framework and \newslink have higher overhead because the queries are performed on a document database instead of search engines. It is possible improve KG based methods by using \lucene to retrieve the matched documents before ranking the results with database records. The implementation is skipped in this study as the sub-second retrieval time does not degrades system usability.
\fi

\subsubsection{Context Relevance Score}\label{sec:csc-effectiveness-study}
\noindent\textbf{Effectiveness and Parameter Study}. 
To verify the effectiveness of the context relevance score $cdr_c(c,d)$ (Eqn.~\ref{eq:conn}), we design a ``negative sampling'' approach.
We randomly select 100 entries from the inverted index $\langle c,d \rangle$.
For each entry $\langle c,d \rangle$, we sample a concept node $c'$ from the KG to generate a \kgentity{negative} concept. 
Fig.~\ref{fig:csc-effectiveness} demonstrates the effective differentiation between $c$ and $c'$ using the context relevance score, i.e., $cdc_c(c,d)>cdc_c(c',d)$, regardless of the hop constraint  $\tau$ from $1$ to $3$.
Notably, when $\tau$ is set to $1$ and $2$, the score differences are significant compared with $\tau=3$, suggesting that a large $\tau$ may include irrelevant concepts due to a higher chance of linking concepts with documents. Our default $\tau$ is set to 2, as over half (55\%) of the relevant scores are 0 when $\tau=1$. In contrast, only 22.4\% of the relevance scores are 0 for $\tau=2$, striking a good balance between information linking and relevance differentiation.

\subsection{Drill-down Operation} \label{sec:subtopic-ranking-study}
\noindent \textbf{Ablation Study.} Given a search query, \framework automatically suggests related concepts for \drilldown operations. \framework ranks the concepts by considering three key factors: \emph{Coverage (C)}, \emph{Specificity (S)} and \emph{Diversity (D)}. To investigate the impact of each factor, we conduct user studies for the ablation analysis. 
We use the same queries from Sec.~\ref{sec:relevance-study} and select top-ranked concepts when only considering: (1) C; (2) C+S; (3) C+S+D.
We build an interactive survey interface (listed in full report) that allows the participants to click on different concepts and view associated results before assigning a distinct rating 1-3 for each subtopic. We recruited participants from the same crowd-source platform AMT~\cite{mturk} and obtain 518 survey results in total. The results are displayed in Fig.~\ref{tab:subtopic_ranking_survey_result}.
We can observe that the specificity contributes slight improvement to the overall rating while diversity plays a more significant role in rating improvement.

\subsection{Roll-up \& Drill-down effectiveness study}
\label{sec:rollup-drilldown-study}

To assess the effectiveness of \rollup and \drilldown in terms of productivity gain at work, we worked with our compliance team to create a task list resembling a standard set of investigative inquiries. The participants are required to investigate issues like: \emph{Find out the \textbf{names of Switzerland Banks} with reports related to \textbf{money laundering}.} 
Given the open-ended nature of these tasks, we set a fixed duration of 2 minutes and use the number of accurate responses as a performance metric. This task design mirrors the requirement of Suspicious Activity Reports (SARs)\cite{sar-report}, where financial institutions must file findings within limited time of identifying potential criminal activity. 
\if\fullversion1
The list of questions are:
\begin{enumerate}[leftmargin=*]
  \item Find out the \textbf{categories of commercial crime} involves \textbf{Credit Suisse}.
  \item Find out the \textbf{categories of commercial crimes} involves \textbf{Banks of Switzerland}.
  \item Find out the \textbf{names of Switzerland Banks} with reports related to \textbf{commercial crimes}.
  \item Find out the \textbf{names of Switzerland Banks} with reports related to \textbf{money laundering}.
  \item Find out the \textbf{categories of commercial crime} involves \textbf{FTX}. 
  \item Find out the \textbf{categories of commercial crimes} involves \textbf{Bitcoin exchanges}.
  \item Find out \textbf{the names of Bitcoin exchanges} with reports related to \textbf{commercial crimes}.
  \item Find out the \textbf{names of Bitcoin exchanges} with reports related to \textbf{money laundering}. 
\end{enumerate}
\else
We have included a detailed description of the study in the extended report\cite{ncexplorerfull} due to space limit. 
\fi

\noindent We recruited 10 financial professionals for this study. Using \framework, professionals can generate more answers within the same time limit than the existing keyword match-based method adopted in the corporation, as shown in Table~\ref{tab:productivity-gain-study}. We also asked a subjective question \kgentity{How likely are you to recommend this tool to due diligence professionals?} and got an average rating of 8.1/10.

\subsection{Connectivity Score Estimation}

\noindent\textbf{RW Estimator Convergence.} 
We evaluate the convergence rate of the proposed RW estimator on the connectivity score. Fig.~\ref{fig:csc-sampling-error} shows the average sampling error of $cdr_c(c,d)$ compared to the ground truth value. Solid and dotted lines represent RW \emph{with} and \emph{without} the guidance of k-hop reachability index respectively. With the k-hop index, our sampling approach can converge on all three datasets within 5\% estimation error using 20 sampling iterations. \\

\if\fullversion1

\begin{figure*}[!htb]
\centering
\begin{adjustwidth}{-1.5cm}{}
      \vspace{-0.5cm}

    \begin{minipage}{.47\linewidth}
        \centering
        \includegraphics[width=\textwidth]{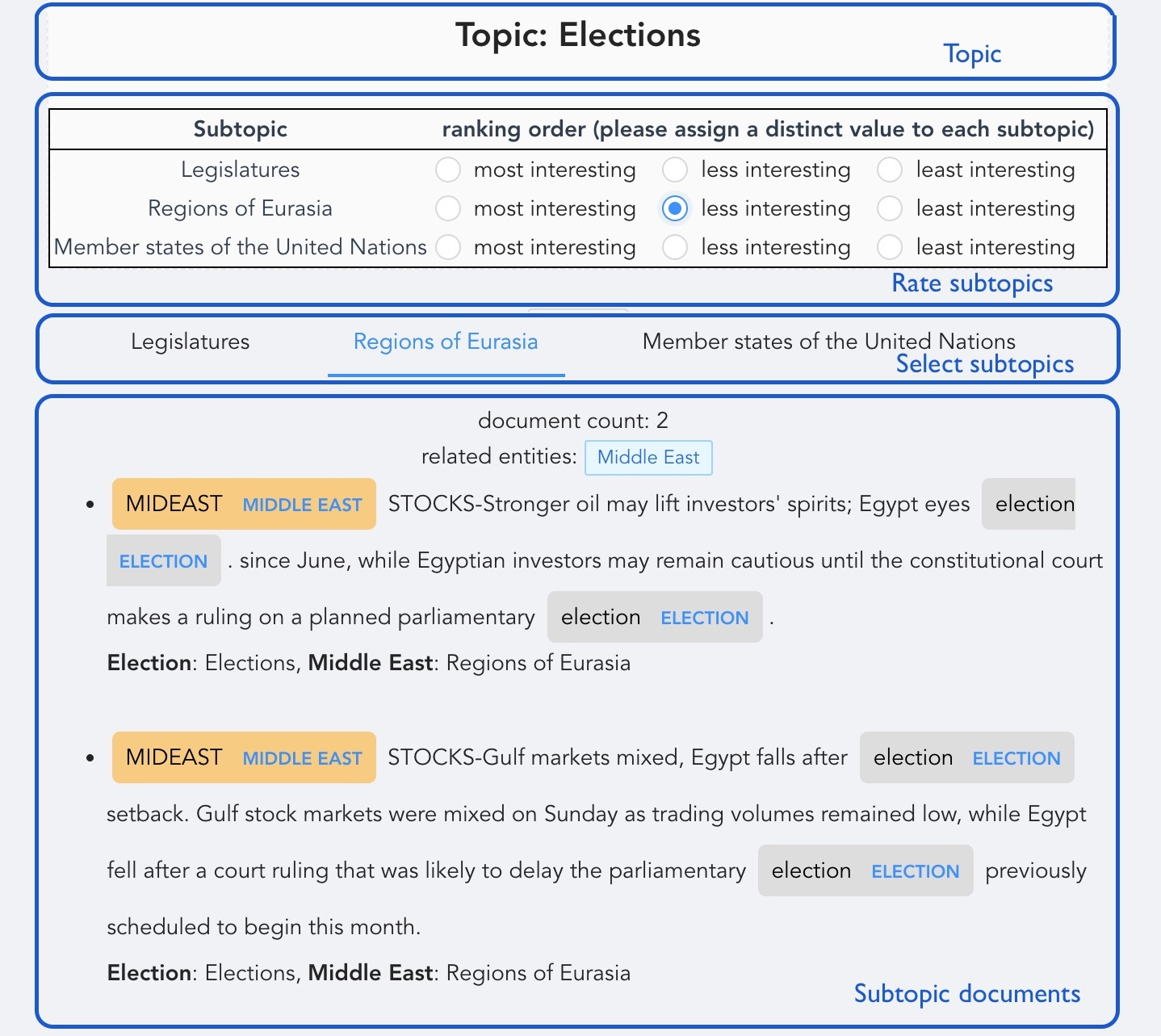}
        \caption{Subtopic survey interface}
        \label{fig:subtopic-survey-interface}
    \end{minipage}
    \hspace{1mm}%
     \begin{minipage}{.63\linewidth}
        \centering
        \includegraphics[width=\textwidth]{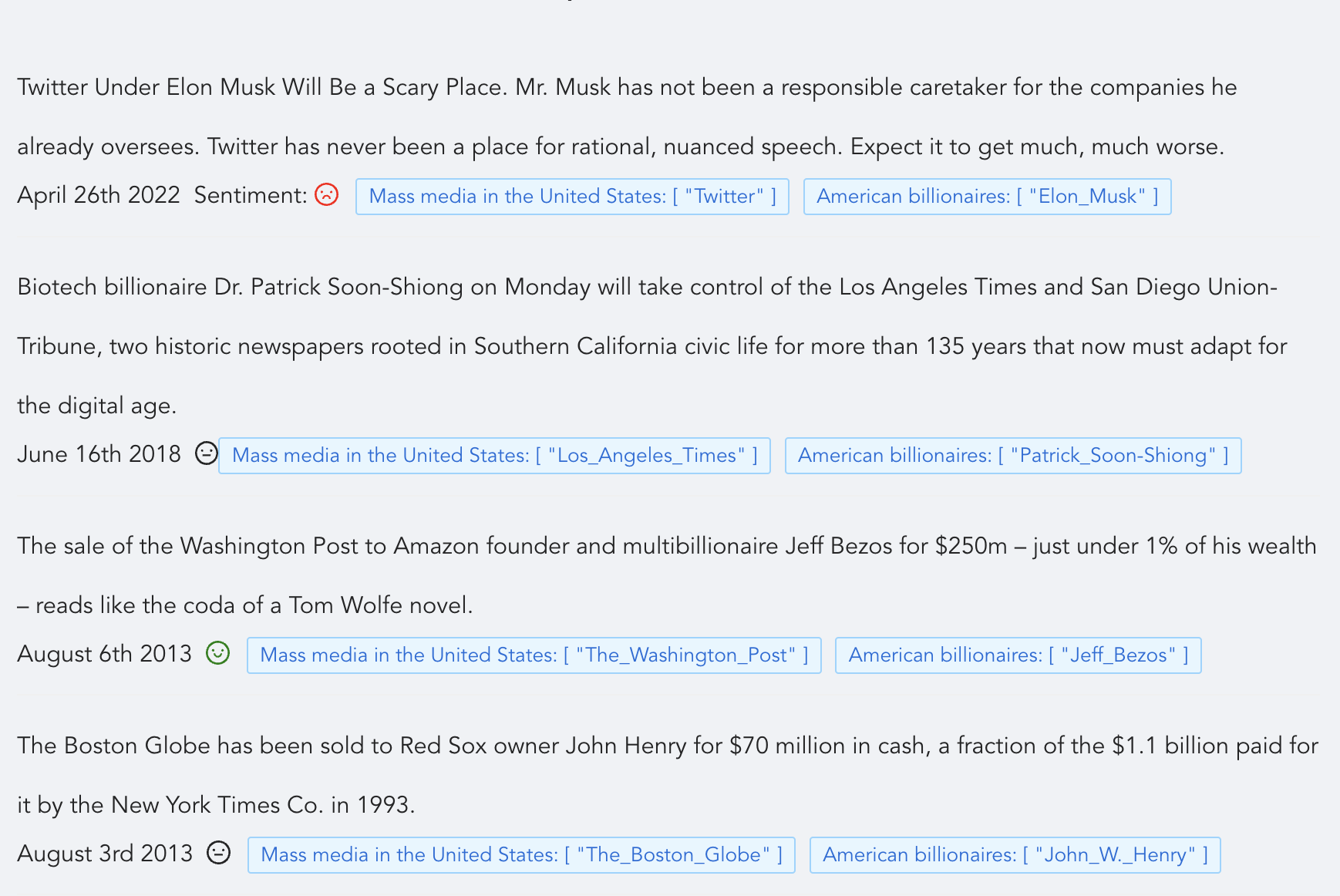}
        \caption{Case study on Media Bias}
        \label{fig:results-with-sentiment}
    \end{minipage}
\end{adjustwidth}
\end{figure*}

\subsection{Case Study on Media Bias}\label{sec:case-study}

In this case study, \framework's capabilities in identifying media bias are demonstrated. When Elon Musk announced his interest in acquiring Twitter, there were intense criticisms towards rich people controlling the media~\cite{twitter-musk} With \framework, users can roll-up \kgentity{Elon Musk} and \kgentity{Twitter} to a concept pattern query \kgentity{American billionaire} and \kgentity{U.S. Mass Media Company}. Among all matched news of the query in the dataset, several similar acquisitions are discovered, such as Jeff Bezos acquiring the Washington Post, Patrick Soon-Shiong buying the Los Angeles Times, and Rupert Murdoch purchasing The Wall Street Journal. Furthermore, the sentiment score of each matched news article is evaluated using a pre-trained model~\cite{perez2021pysentimiento}. Interestingly, Musk's acquisition is the only news with a negative sentiment. The rest have either neutral or positive sentiments (Fig~\ref{fig:results-with-sentiment}). Such insights reveal a potential news bias that deserves further investigation by experts. This example showcases how \framework can be a powerful tool for uncovering hidden patterns and biases in news coverage.

\fi
\section{Conclusion}
\label{chap:conclusion}
\framework is designed to enhance the news article exploration experience by using OLAP-like operations to connect them with relevant concepts in a KG.      \framework not only streamlines due diligence tasks but also improves a range of news analytics tasks. Its efficacy has been validated through crowd-sourced evaluations with a dataset of real-world news articles and a large knowledge graph. Its modular design ensures compatibility with various text-based systems, such as search engines and literature databases, facilitating information discovery and comprehension. Additionally, the release of 200,000 news articles, along with their KG annotations and concept relevance scores, allows for easy customization of the tool to meet specific research needs.


%
%
%

\bibliographystyle{IEEEtran}
\bibliography{main}

\begin{thebibliography}{10}
\providecommand{\url}[1]{#1}
\csname url@samestyle\endcsname
\providecommand{\newblock}{\relax}
\providecommand{\bibinfo}[2]{#2}
\providecommand{\BIBentrySTDinterwordspacing}{\spaceskip=0pt\relax}
\providecommand{\BIBentryALTinterwordstretchfactor}{4}
\providecommand{\BIBentryALTinterwordspacing}{\spaceskip=\fontdimen2\font plus
\BIBentryALTinterwordstretchfactor\fontdimen3\font minus \fontdimen4\font\relax}
\providecommand{\BIBforeignlanguage}[2]{{%
\expandafter\ifx\csname l@#1\endcsname\relax
\typeout{** WARNING: IEEEtran.bst: No hyphenation pattern has been}%
\typeout{** loaded for the language `#1'. Using the pattern for}%
\typeout{** the default language instead.}%
\else
\language=\csname l@#1\endcsname
\fi
#2}}
\providecommand{\BIBdecl}{\relax}
\BIBdecl

\bibitem{compliance-commitment-paypal}
\BIBentryALTinterwordspacing
Paypal: Anti-money laundering and know your customer. Accessed: 2023-10-01. [Online]. Available: \url{https://publicpolicy.paypal-corp.com/issues/anti-money-laundering-know-your-customer}
\BIBentrySTDinterwordspacing

\bibitem{dbs-responsible-finance}
\BIBentryALTinterwordspacing
(2021, Apr.) Dbs bank: Our approach to responsible financing. Accessed: 2023-10-01. [Online]. Available: \url{https://www.dbs.com/iwov-resources/images/sustainability/responsible-banking/DBS%20Bank_Our%20Approach%20to%20Responsible%20Financing_Updated_26Apr2021.pdf}
\BIBentrySTDinterwordspacing

\bibitem{compliance-cost-mckinsey}
\BIBentryALTinterwordspacing
(2019) The compliance function at an inflection point. [Online]. Available: \url{https://www.mckinsey.com/capabilities/risk-and-resilience/our-insights/the-compliance-function-at-an-inflection-point}
\BIBentrySTDinterwordspacing

\bibitem{twitter-musk}
\BIBentryALTinterwordspacing
G.~Bensinger. (2022) Twitter under elon musk will be a scary place. Accessed: 2023-05-22. [Online]. Available: \url{https://www.nytimes.com/2022/04/25/opinion/editorials/twitter-elon-musk.html}
\BIBentrySTDinterwordspacing

\bibitem{bezo-washington-post}
\BIBentryALTinterwordspacing
E.~Bell. (2013) Jeff bezos' shocking washington post buy was not a business deal — it was a cultural statement. Accessed: 2023-05-02. [Online]. Available: \url{https://twitter.com/BusinessInsider/status/364551288845254656}
\BIBentrySTDinterwordspacing

\bibitem{patrick-la-times}
\BIBentryALTinterwordspacing
J.~R.~K. Meg~James. (2018) Billionaire patrick soon-shiong reaches deal to buy l.a. times and san diego union-tribune. Accessed: 2023-05-02. [Online]. Available: \url{https://www.latimes.com/business/hollywood/la-fi-ct-los-angeles-times-sold-20180207-story.html}
\BIBentrySTDinterwordspacing

\bibitem{murdoch-dow-jones}
\BIBentryALTinterwordspacing
M.~K. Sarah~Ellison. (2007) Murdoch wins his bid for dow jones. Accessed: 2023-05-02. [Online]. Available: \url{https://www.wsj.com/articles/SB118589043953483378}
\BIBentrySTDinterwordspacing

\bibitem{tanon2020yago}
T.~P. Tanon, G.~Weikum, and F.~Suchanek, ``Yago 4: A reason-able knowledge base,'' in \emph{ESWC}, 2020, pp. 583--596.

\bibitem{lehmann2015dbpedia}
e.~a. Lehmann, ``Dbpedia, a large-scale, multilingual knowledge base extracted from wikipedia,'' \emph{Semantic web}, 2015.

\bibitem{ncexplorersource}
\BIBentryALTinterwordspacing
Ncexplorer source code. [Online]. Available: \url{https://github.com/knowledge-fusion/ncexplorer/}
\BIBentrySTDinterwordspacing

\bibitem{dbpedia2016snapshot}
\BIBentryALTinterwordspacing
Dbpedia snapshot 2021-06 release. Accessed: 2023-02-22. [Online]. Available: \url{https://www.dbpedia.org/blog/snapshot-2021-06-release/}
\BIBentrySTDinterwordspacing

\bibitem{devlin2018bert}
J.~Devlin, M.-W. Chang, K.~Lee, and K.~Toutanova, ``Bert: Pre-training of deep bidirectional transformers for language understanding,'' \emph{arXiv preprint arXiv:1810.04805}, 2018.

\bibitem{reimers2019sentence}
N.~Reimers and I.~Gurevych, ``Sentence-bert: Sentence embeddings using siamese bert-networks,'' \emph{arXiv preprint arXiv:1908.10084}, 2019.

\bibitem{sheu2020context}
H.-S. Sheu and S.~Li, ``Context-aware graph embedding for session-based news recommendation,'' in \emph{RecSys}, 2020, pp. 657--662.

\bibitem{wang2018dkn}
H.~Wang, F.~Zhang, X.~Xie, and M.~Guo, ``Dkn: Deep knowledge-aware network for news recommendation,'' in \emph{Proceedings of the 2018 world wide web conference}, 2018, pp. 1835--1844.

\bibitem{liu2020kred}
D.~Liu, J.~Lian, S.~Wang, Y.~Qiao, J.-H. Chen, G.~Sun, and X.~Xie, ``Kred: Knowledge-aware document representation for news recommendations,'' in \emph{Proceedings of the 14th ACM Conference on Recommender Systems}, 2020, pp. 200--209.

\bibitem{newsgraph}
D.~Liu, T.~Bai, J.~Lian, X.~Zhao, G.~Sun, J.-R. Wen, and X.~Xie, ``News graph: An enhanced knowledge graph for news recommendation.'' in \emph{KaRS@ CIKM}, 2019, pp. 1--7.

\bibitem{kim2018deep}
J.~Kim, A.-D. Nguyen, and S.~Lee, ``Deep cnn-based blind image quality predictor,'' \emph{IEEE transactions on neural networks and learning systems}, vol.~30, no.~1, pp. 11--24, 2018.

\bibitem{transd}
G.~Ji, S.~He, L.~Xu, K.~Liu, and J.~Zhao, ``Knowledge graph embedding via dynamic mapping matrix,'' in \emph{Proceedings of the 53rd annual meeting of the association for computational linguistics and the 7th international joint conference on natural language processing (volume 1: Long papers)}, 2015, pp. 687--696.

\bibitem{DBLP:conf/kdd/LiuLLWS021}
D.~Liu, J.~Lian, Z.~Liu, X.~Wang, G.~Sun, and X.~Xie, ``Reinforced anchor knowledge graph generation for news recommendation reasoning,'' in \emph{KDD}.\hskip 1em plus 0.5em minus 0.4em\relax {ACM}, 2021, pp. 1055--1065.

\bibitem{yang2021newslink}
Y.~Yang, Y.~Li, and A.~K. Tung, ``Newslink: Empowering intuitive news search with knowledge graphs,'' in \emph{ICDE}, 2021, pp. 876--887.

\bibitem{fan2017discovering}
Q.~Fan, Y.~Li, D.~Zhang, and K.-L. Tan, ``Discovering newsworthy themes from sequenced data: A step towards computational journalism,'' \emph{IEEE TKDE}, vol.~29, no.~7, pp. 1398--1411, 2017.

\bibitem{yang2021context}
Y.~Yang, Y.~Li, P.~Karras, and A.~K. Tung, ``Context-aware outstanding fact mining from knowledge graphs,'' in \emph{Proceedings of the 27th ACM SIGKDD Conference on Knowledge Discovery \& Data Mining}, 2021, pp. 2006--2016.

\bibitem{honnibal2017spacy}
M.~Honnibal and I.~Montani, ``spacy 2: Natural language understanding with bloom embeddings, convolutional neural networks and incremental parsing,'' \emph{To appear}, vol.~7, no.~1, pp. 411--420, 2017.

\bibitem{qin2019towards}
L.~Qin, Y.~Peng, Y.~Zhang, X.~Lin, W.~Zhang, and J.~Zhou, ``Towards bridging theory and practice: hop-constrained st simple path enumeration,'' in \emph{International Conference on Very Large Data Bases}.\hskip 1em plus 0.5em minus 0.4em\relax VLDB Endowment, 2019.

\bibitem{sun2021pathenum}
S.~Sun, Y.~Chen, B.~He, and B.~Hooi, ``Pathenum: Towards real-time hop-constrained st path enumeration,'' in \emph{SIGMOD}, 2021, pp. 1758--1770.

\bibitem{li2016wander}
F.~Li, B.~Wu, K.~Yi, and Z.~Zhao, ``Wander join: Online aggregation via random walks,'' in \emph{SIGMOD}, 2016, pp. 615--629.

\bibitem{zhao2018random}
Z.~Zhao, R.~Christensen, F.~Li, X.~Hu, and K.~Yi, ``Random sampling over joins revisited,'' in \emph{SIGMOD}, 2018, pp. 1525--1539.

\bibitem{park2020g}
Y.~Park, S.~Ko, S.~S. Bhowmick, K.~Kim, K.~Hong, and W.-S. Han, ``G-care: a framework for performance benchmarking of cardinality estimation techniques for subgraph matching,'' in \emph{SIGMOD}, 2020, pp. 1099--1114.

\bibitem{sun2021thunderrw}
S.~Sun, Y.~Chen, S.~Lu, B.~He, and Y.~Li, ``Thunderrw: an in-memory graph random walk engine,'' \emph{PVLDB}, vol.~14, no.~11, pp. 1992--2005, 2021.

\bibitem{cheng2014efficient}
J.~Cheng, Z.~Shang, H.~Cheng, H.~Wang, and J.~X. Yu, ``Efficient processing of k-hop reachability queries,'' \emph{The VLDB journal}, vol.~23, no.~2, pp. 227--252, 2014.

\bibitem{ncexplorerfull}
\BIBentryALTinterwordspacing
``Ncexplorer full report,'' Tech. Rep. [Online]. Available: \url{https://github.com/knowledge-fusion/ncexplorer/blob/github-main/NCExplorer_full_report.pdf}
\BIBentrySTDinterwordspacing

\bibitem{reuters}
\BIBentryALTinterwordspacing
Reuters. Accessed: 2023-02-22. [Online]. Available: \url{https://www.reuters.com}
\BIBentrySTDinterwordspacing

\bibitem{seekingalpha}
\BIBentryALTinterwordspacing
Seeking alpha. Accessed: 2023-02-22. [Online]. Available: \url{https://seekingalpha.com}
\BIBentrySTDinterwordspacing

\bibitem{nytimes}
\BIBentryALTinterwordspacing
The new york times. Accessed: 2023-02-22. [Online]. Available: \url{https://www.nytimes.com}
\BIBentrySTDinterwordspacing

\bibitem{robertson2009probabilistic}
S.~Robertson, H.~Zaragoza \emph{et~al.}, ``The probabilistic relevance framework: Bm25 and beyond,'' \emph{Foundations and Trends{\textregistered} in Information Retrieval}, vol.~3, no.~4, pp. 333--389, 2009.

\bibitem{sbert-model}
\BIBentryALTinterwordspacing
Sentence transformers all-mpnet-base-v2. [Online]. Available: \url{https://huggingface.co/sentence-transformers/all-mpnet-base-v2}
\BIBentrySTDinterwordspacing

\bibitem{qdrant}
\BIBentryALTinterwordspacing
``Vector search engine.'' [Online]. Available: \url{https://qdrant.tech/}
\BIBentrySTDinterwordspacing

\bibitem{gpt-3.5-turbo}
\BIBentryALTinterwordspacing
Openai models gpt-3.5. [Online]. Available: \url{https://platform.openai.com/docs/models/gpt-3-5}
\BIBentrySTDinterwordspacing

\bibitem{gpt-4}
\BIBentryALTinterwordspacing
Openai models gpt-4. [Online]. Available: \url{https://platform.openai.com/docs/models/gpt-4}
\BIBentrySTDinterwordspacing

\bibitem{bloomberg-stock-futures}
\BIBentryALTinterwordspacing
Bloomberg. Market futures. Accessed: 2023-05-02. [Online]. Available: \url{https://www.bloomberg.com/markets/stocks/futures}
\BIBentrySTDinterwordspacing

\bibitem{mturk}
\BIBentryALTinterwordspacing
I.~Amazon Mechanical~Turk. Amazon mechanical turk. [Online]. Available: \url{https://www.mturk.com}
\BIBentrySTDinterwordspacing

\bibitem{sar-report}
\BIBentryALTinterwordspacing
T.~O. of~the Comptroller of~the Currency. Suspicious activity reports (sar). Accessed: 2024-02-02. [Online]. Available: \url{https://www.occ.treas.gov/topics/supervision-and-examination/bank-operations/financial-crime/suspicious-activity-reports/index-suspicious-activity-reports.html}
\BIBentrySTDinterwordspacing

\end{thebibliography}


%

\end{document}